\documentclass[a4paper, oneside, twocolumn, notitlepage, 10pt]{extarticle_ecoc}
\usepackage{ecoc}
\usepackage[helvet]{sfmath}
\usepackage[binary-units, per-mode=symbol]{siunitx}
\DeclareSIUnit{\sample}{Sa}
\DeclareSIUnit{\baud}{Bd}
\DeclareSIUnit{\bit}{bit}
\DeclareSIUnit{\fourd}{4D}
\DeclareSIUnit{\eightd}{8D}
\DeclareSIUnit{\dBm}{dBm}
\DeclareSIUnit{\dB}{dB}
\DeclareSIUnit{\bps}{bps}

\usepackage{glossaries}
\glsdisablehyper
\newcommand{\SetCapsType}{normalcaps}

\makeatletter
\providecommand{\SetCapsType}{smallcaps}

\long\def\@scTrue{smallcaps}
\long\def\@scFalse{normalcaps}
\newcommand{\acroSCaps}[1]{%
 \begingroup
  \ifx\SetCapsType\@scTrue 
    \textsc{#1}%
  \else
    \MakeUppercase{#1}%
  \fi
  \endgroup
}
\makeatother

\newcommand{\nAcronym}[4][]{%
	\newacronym[#1]{#2}{#3}{#4}
}

\makeatletter
\@ifpackageloaded{babel}{%
    \newcommand{\usuk}[2]{%
        \iflanguage{USenglish}{#1}{#2}%
    }%
}{%
    \newcommand{\usuk}[2]{%
        #1%
    }%
}%
\makeatother

\usepackage[shortcuts]{extdash}
\nAcronym{2A8PSK}{\acroSCaps{2a8psk}}{2-ary amplitude 8-ary phaseshift keying}

\nAcronym{3CCMCF}{\acroSCaps{3cc-mcf}}{3-core coupled-core multi-core fiber}

\nAcronym{4D}{\acroSCaps{4d}}{four-dimensional}
\nAcronym{4D64PRS}{\acroSCaps{4d-64prs}}{\usuk{four-dimensional 64-ary polarization-ring-switching}{four-dimensional 64-ary polarisation-ring-switching}}

\nAcronym{5B4D2A8PSK}{\acroSCaps{5b4d-2a8psk}}{5-bit four-dimensional two-amplitude 8-ary phase-shift keying}

\nAcronym{6B4D2A8PSK}{\acroSCaps{6b4d-2a8psk}}{6-bit four-dimensional two-amplitude 8-ary phase-shift keying}

\nAcronym{8D}{\acroSCaps{8d}}{eight-dimensional}\nAcronym{8D2048PRS}{\acroSCaps{8d-2048prs}}{eight-dimensional 2048-ary polarization-ring-switching}
\nAcronym{8D2048PRST1}{\acroSCaps{8d-2048prs-t1}}{eight-dimensional 2048-ary polarization-ring-switching type 1}
\nAcronym{8D2048PRST2}{\acroSCaps{8d-2048prs-t2}}{eight-dimensional 2048-ary polarization-ring-switching type 2}

\nAcronym{ABC}{\acroSCaps{abc}}{automatic bias controller}
\nAcronym{ADC}{\acroSCaps{adc}}{analog-to-digital converter}
\nAcronym{AIR}{\acroSCaps{air}}{achievable information rate}
\nAcronym{AOM}{\acroSCaps{aom}}{acoustic optical modulator}
\nAcronym{API}{\acroSCaps{api}}{application programming interface}
\nAcronym{AR}{\acroSCaps{ar}}{achievable rate}
\nAcronym{ASE}{\acroSCaps{ase}}{amplified spontaneous emission}
\nAcronym{ASK}{\acroSCaps{ask}}{amplitude-shift keying}
\nAcronym{ASIC}{\acroSCaps{asic}}{application-specific integrated circuit}
\nAcronym{AWGN}{\acroSCaps{awgn}}{additive white Gaussian noise}

\nAcronym{BCH}{\acroSCaps{bch}}{Bose-Chaudhuri-Hocquenghem}
\nAcronym{BER}{\acroSCaps{ber}}{bit error rate}
\nAcronym{BICM}{\acroSCaps{bicm}}{bit-interleaved coded modulation}
\nAcronym{BMD}{\acroSCaps{bmd}}{bit-metric decoding}
\nAcronym{BPD}{\acroSCaps{bpd}}{balanced photo-diode}
\nAcronym{BPS}{\acroSCaps{bps}}{blind phase search}
\nAcronym{BPSK}{\acroSCaps{bpsk}}{binary phase-shift keying}
\nAcronym{BRGC}{\acroSCaps{brgc}}{binary reflected Gray code}
\nAcronym{BTB}{\acroSCaps{btb}}{back-to-back}

\nAcronym{CAGR}{\acroSCaps{cagr}}{compound annual growth rate}
\nAcronym{CCDM}{\acroSCaps{ccdm}}{constant composition distribution matching}
\nAcronym{CCF}{\acroSCaps{ccf}}{\usuk{coupled-core fiber}{coupled-core fibre}}
\nAcronym{CD}{\acroSCaps{cd}}{chromatic dispersion}
\nAcronym{CIR}{\acroSCaps{cir}}{channel impulse response}
\nAcronym{CMA}{\acroSCaps{cma}}{constant modulus algorithm}
\nAcronym{CMUX}{\acroSCaps{cmux}}{core multiplexer}
\nAcronym{ChUT}{\acroSCaps{chut}}{channel under test}
\nAcronym{CUT}{\acroSCaps{cut}}{channel under test}
\nAcronym{CPE}{\acroSCaps{cpe}}{carrier phase estimation}
\nAcronym{CPU}{\acroSCaps{cpu}}{central processing unit}
\nAcronym{CW}{\acroSCaps{cw}}{continuous wave}

\nAcronym{DA}{\acroSCaps{da}}{driver amplifier}
\nAcronym{DAC}{\acroSCaps{dac}}{digital-to-analog converter}
\nAcronym{DCF}{\acroSCaps{dcf}}{\usuk{dispersion compensated fiber}{dispersion compensated fibre}}
\nAcronym{DCI}{\acroSCaps{dci}}{data center interconnect}
\nAcronym{DDLMS}{\acroSCaps{dd-lms}}{decision-directed least mean square}
\nAcronym{DFB}{\acroSCaps{dfb}}{distributed feedback}
\nAcronym{DGD}{\acroSCaps{dgd}}{differential group delay}
\nAcronym{DH}{\acroSCaps{dh}}{digital holography}
\nAcronym{DM}{\acroSCaps{dm}}{distribution matcher}
\nAcronym{DMA}{\acroSCaps{dma}}{direct memory access}
\nAcronym{DMD}{\acroSCaps{dmd}}{differential mode delay}
\nAcronym{DMG}{\acroSCaps{dmg}}{differential modal gain}
\nAcronym{DMGD}{\acroSCaps{dmgd}}{differential mode group delay}
\nAcronym{DML}{\acroSCaps{dml}}{directly-modulated laser}
\nAcronym{DP}{\acroSCaps{dp}}{\usuk{dual-polarization}{dual-polarisation}}
\nAcronym{DPC}{\acroSCaps{dpc}}{digital pre-compensation}
\nAcronym{DPE}{\acroSCaps{dpe}}{digital pre-emphasis}
\nAcronym{DPIQ}{\acroSCaps{dp-iqm}}{\usuk{dual-polarization IQ-modulator}{dual-polarisation IQ-modulator}}
\nAcronym{DRA}{\acroSCaps{dra}}{distributed Raman amplifier}
\nAcronym{DRE}{\acroSCaps{dre}}{digital resolution enhancer}
\nAcronym{DSF}{\acroSCaps{dsf}}{\usuk{dispersion-shifted fiber}{dispersion-shifted fibre}}
\nAcronym{DSO}{\acroSCaps{dso}}{digital sampling oscilloscope}
\nAcronym{DSP}{\acroSCaps{dsp}}{digital signal processing}
\nAcronym{DPLL}{\acroSCaps{dpll}}{digital phase-locked loop}
\nAcronym{DUT}{\acroSCaps{dut}}{device-under-test}
\nAcronym{DWDM}{\acroSCaps{dwdm}}{dense wavelength-division multiplexing}

\nAcronym{EAM}{\acroSCaps{eam}}{electro-absorption modulator}
\nAcronym{ECL}{\acroSCaps{ecl}}{external cavity laser}
\nAcronym{ED}{\acroSCaps{ed}}{Eucledian distance}
\nAcronym{EDFA}{\acroSCaps{edfa}}{\usuk{erbium-doped fiber amplifier}{erbium-doped fibre amplifier}}
\nAcronym{ENOB}{\acroSCaps{enob}}{effective number of bits}
\nAcronym{ESS}{\acroSCaps{ess}}{enumerative sphere shaping}

\nAcronym{FD}{\acroSCaps{fd}}{frequency domain}
\nAcronym{FDE}{\acroSCaps{fde}}{\usuk{frequency domain equalizer}{frequency domain equaliser}}
\nAcronym{FEC}{\acroSCaps{fec}}{forward error correction}
\nAcronym{FFT}{\acroSCaps{fft}}{fast Fourier transform}
\nAcronym{FIR}{\acroSCaps{fir}}{finite impulse response}
\nAcronym{FMF}{\acroSCaps{fmf}}{\usuk{few-mode fiber}{few-mode fibre}}
\nAcronym[plural=FM-MCF, firstplural=\usuk{few-mode multi-core fibers}{few-mode multi-core fibres}]{FM-MCF}{\acroSCaps{fm-mcf}}{\usuk{few-mode multi-core fiber}{few-mode multi-core fibre}}
\nAcronym{FPGA}{\acroSCaps{fpga}}{field-programmable gate array}
\nAcronym{FWM}{\acroSCaps{fwm}}{four-wave mixing}

\nAcronym{GD}{\acroSCaps{gd}}{group delay}
\nAcronym{GI}{\acroSCaps{gi}}{graded-index}
\nAcronym{GFF}{\acroSCaps{gff}}{gain flattening filter}
\nAcronym{GIMMF}{\acroSCaps{gi-mmf}}{\usuk{graded-index multi-mode fiber}{graded-index multi-mode fibre}}
\nAcronym{GMI}{\acroSCaps{gmi}}{\usuk{generalized mutual information}{generalised mutual information}}
\nAcronym{GNSE}{\acroSCaps{gnse}}{generalized nonlinear Schr\"{o}dinger equation}
\nAcronym{GPU}{\acroSCaps{gpu}}{graphics processing unit}
\nAcronym{GS}{\acroSCaps{gs}}{geometric shaping}
\nAcronym{GV}{\acroSCaps{gv}}{group-velocity}
\nAcronym{GVD}{\acroSCaps{gvd}}{group-velocity dispersion}
\nAcronym{GPIO}{\acroSCaps{gpio}}{general purpose input output}

\nAcronym{HDFEC}{\acroSCaps{hd-fec}}{hard decision forward error correction}

\nAcronym[plural=IL, firstplural=insertion losses (\textsc{il})]{IL}{\acroSCaps{il}}{insertion loss}
\nAcronym{IFFT}{\acroSCaps{ifft}}{inverse fast Fourier transform}
\nAcronym{IMDD}{\acroSCaps{im/dd}}{intensity-modulation direct-detection}
\nAcronym{ISI}{\acroSCaps{isi}}{intersymbol interference}

\nAcronym{KK}{\acroSCaps{kk}}{Kramers-Kronig}

\nAcronym{LCOS}{\acroSCaps{LCoS}}{liquid crystal on silicon}
\nAcronym{LDPC}{\acroSCaps{ldpc}}{low-density parity-check}
\nAcronym{LEAF}{\acroSCaps{leaf}}{\usuk{large effective area fiber}{large effective area fibre}}
\nAcronym{LFSR}{\acroSCaps{lfsr}}{linear-feedback shift register}
\nAcronym{LMS}{\acroSCaps{lmf}}{least means square}
\nAcronym{LLR}{\acroSCaps{llr}}{log-likelihood ratio}
\nAcronym{LO}{\acroSCaps{lo}}{local oscillator}
\nAcronym{LP}{\acroSCaps{lp}}{\usuk{linear polarized}{linear polarised}}
\nAcronym{LSPS}{\acroSCaps{lsps}}{\usuk{loop-synchronized polarization scrambler}{loop-synchronised polarisation scrambler}}
\nAcronym{LUT}{\acroSCaps{lut}}{lookup table}

\nAcronym{MB}{\acroSCaps{mb}}{Maxwell-Bolzmann}
\nAcronym{MCF}{\acroSCaps{mcf}}{\usuk{multi-core fiber}{multi-core fibre}}
\nAcronym{MDG}{\acroSCaps{mdg}}{mode dependent gain}
\nAcronym[plural=MDL, firstplural=mode-dependent losses (\textsc{mdl})]{MDL}{\acroSCaps{mdl}}{mode-dependent loss}
\nAcronym{MDM}{\acroSCaps{mdm}}{mode division multiplexing}
\nAcronym{MEMS}{\acroSCaps{mems}}{micro-electro-mechanical systems}
\nAcronym{MF}{\acroSCaps{mf}}{matched filter}
\nAcronym{MI}{\acroSCaps{mi}}{mutual information}
\nAcronym{MIMO}{\acroSCaps{mimo}}{multiple-input multiple-output}
\nAcronym{MMA}{\acroSCaps{mma}}{multi-modulus algorithm}
\nAcronym{MMF}{\acroSCaps{mmf}}{\usuk{multi-mode fiber}{multi-mode fibre}}
\nAcronym{MMSE}{\acroSCaps{mmse}}{minimum mean squared error}
\nAcronym{MPLC}{\acroSCaps{mplc}}{multi-plane light converter}
\nAcronym{MSE}{\acroSCaps{mse}}{mean squared error}
\nAcronym{MUX}{\acroSCaps{mux}}{multiplexer}
\nAcronym{MZM}{\acroSCaps{mzm}}{Mach-Zehnder modulator}
\nAcronym{MZI}{\acroSCaps{mzi}}{Mach-Zehnder interferometer}

\nAcronym{NA}{\acroSCaps{na}}{numerical aperture}
\nAcronym{NF}{\acroSCaps{nf}}{noise figure}
\nAcronym{NGMI}{\acroSCaps{ngmi}}{\usuk{normalized generalized mutual information}{normalised generalised mutual information}}
\nAcronym{NLSE}{\acroSCaps{nlse}}{nonlinear Schr\"{o}ding equation}
\nAcronym{NIR}{\acroSCaps{nir}}{near infra-red}

\nAcronym{OBTB}{\acroSCaps{obtb}}{optical back-to-back}
\nAcronym{ODE}{\acroSCaps{ode}}{ordinary differential equation}
\nAcronym{ODL}{\acroSCaps{odl}}{optical delay line}
\nAcronym{OFDR}{\acroSCaps{ofdr}}{optical frequency-domain reflectometer}
\nAcronym{OFDM}{\acroSCaps{ofdm}}{orthogonal frequency division multiplexing}
\nAcronym{OMFT}{\acroSCaps{omft}}{optical multi-format transmitter}
\nAcronym{OOK}{\acroSCaps{ook}}{on-off keying}
\nAcronym{OPLL}{\acroSCaps{opll}}{optical phase-locked loop}
\nAcronym{OSA}{\acroSCaps{osa}}{\usuk{optical spectrum analyzer}{optical spectrum analyser}}
\nAcronym{OSNR}{\acroSCaps{osnr}}{optical signal-to-noise ratio}
\nAcronym{OTDR}{\acroSCaps{otdr}}{optical time-domain reflectometer}
\nAcronym{OTF}{\acroSCaps{otf}}{optical tunable filter}
\nAcronym{OVNA}{\acroSCaps{ovna}}{\usuk{optical vector network analyzer}{optical vector network analyser}}

\nAcronym{PAM}{\acroSCaps{pam}}{pulse-amplitude modulation}
\nAcronym{PAS}{\acroSCaps{pas}}{probabilistic amplitude shaping}
\nAcronym{PAPR}{\acroSCaps{papr}}{peak-to-avarage power ratio}
\nAcronym{PBC}{\acroSCaps{pbc}}{\usuk{polarization beam combiner}{polarisation beam combiner}}
\nAcronym{PBS}{\acroSCaps{pbs}}{polarization beam splitter}
\nAcronym{PCVD}{\acroSCaps{pcvd}}{plasma chemical vapor depostion}
\nAcronym{PD}{\acroSCaps{pd}}{photodiode}
\nAcronym{PDL}{\acroSCaps{pdl}}{polarization-dependent loss}
\nAcronym{PDM}{\acroSCaps{pdm}}{\usuk{polarization-division multiplexing}{polarisation-division multiplexing}}
\nAcronym{PL}{\acroSCaps{pl}}{photonic lantern}
\nAcronym{PMBPSK}{\acroSCaps{pm-bpsk}}{polarization-multiplexed binary phase-shift-keying}
\nAcronym{PMQPSK}{\acroSCaps{pm-qpsk}}{polarization-multiplexed quaternary phase-shift-keying}
\nAcronym{PM8QAM}{\acroSCaps{pm-8qam}}{polarization-multiplexed 8-ary quadrature amplitude modulation}
\nAcronym{PMD}{\acroSCaps{pmd}}{polarization mode dispersion}
\nAcronym{PNOB}{\acroSCaps{pnob}}{physical number of bits}
\nAcronym{PON}{\acroSCaps{pon}}{passive-optical network}
\nAcronym{PRBS}{\acroSCaps{prbs}}{pseudorandom bit sequence}
\nAcronym{PS}{\acroSCaps{ps}}{probabilistic shaping}
\nAcronym{PSK}{\acroSCaps{psk}}{phase-shift keying}
\nAcronym{PSP}{\acroSCaps{psp}}{\usuk{principal states of polarization}{principal states of polarisation}}

\nAcronym{QAM}{\acroSCaps{qam}}{quadrature amplitude modulation}
\nAcronym{QPSK}{\acroSCaps{qpsk}}{quaternary phase-shift-keying}

\nAcronym{RAM}{\acroSCaps{ram}}{random-access memory}
\nAcronym{RF}{\acroSCaps{rf}}{radio frequency}
\nAcronym{RRC}{\acroSCaps{rrc}}{root-raised-cosine}
\nAcronym{ROADM}{\acroSCaps{roadm}}{reconfigurable optical add-drop multiplexer}
\nAcronym{ROI}{\acroSCaps{roi}}{region of interest}

\nAcronym{SA}{\acroSCaps{sa}}{simulated annealing}
\nAcronym{SamPerSym}{\acroSCaps{sps}}{samples per symbol}
\nAcronym{SBS}{\acroSCaps{sbs}}{stimulated Brillouin scattering}
\nAcronym{SDFEC}{\acroSCaps{sd-fec}}{soft decision forward error correction}
\nAcronym{SDM}{\acroSCaps{sdm}}{space-division multiplexing}
\nAcronym{SE}{\acroSCaps{se}}{spectral efficiency}
\nAcronym{SER}{\acroSCaps{ser}}{symbol error rate}
\nAcronym{SI}{\acroSCaps{si}}{step index}
\nAcronym{SLM}{\acroSCaps{slm}}{spatial light modulator}
\nAcronym{SMF}{\acroSCaps{smf}}{\usuk{single-mode fiber}{single-mode fibre}}
\nAcronym{SNR}{\acroSCaps{snr}}{signal-to-noise ratio}
\nAcronym{SOA}{\acroSCaps{soa}}{semiconductor optical amplifier}
\nAcronym[firstplural=\usuk{states of polarization (\acroSCaps{sop})}{states of polarisation (\acroSCaps{sop})}]{SOP}{\acroSCaps{sop}}{\usuk{state of polarization}{state of polarisation}}
\nAcronym{SPM}{\acroSCaps{spm}}{self-phase modulation}
\nAcronym{SPS}{\acroSCaps{sps}}{\usuk{synchronized polarization scrambler}{synchronised polarisation scrambler}}
\nAcronym{SRS}{\acroSCaps{srs}}{stimulated Raman scattering}
\nAcronym{SSFM}{\acroSCaps{ssfm}}{split-step Fourier method}
\nAcronym{SSMF}{\acroSCaps{ssmf}}{\usuk{standard single-mode fiber}{standard single-mode fibre}}
\nAcronym{STL}{\acroSCaps{stl}}{swept tunable laser}
\nAcronym{SVD}{\acroSCaps{svd}}{singular value decomposition}
\nAcronym{SWI}{\acroSCaps{swi}}{swept wavelength interferometry}

\nAcronym{TD}{\acroSCaps{td}}{time domain}
\nAcronym{TDE}{\acroSCaps{tde}}{\usuk{time domain equalizer}{time domain equaliser}}
\nAcronym{TDMSDM}{\acroSCaps{tdm-sdm}}{time-domain multiplexed space-division multiplexing}
\nAcronym{TE}{\acroSCaps{te}}{transverse electric}
\nAcronym{TIA}{\acroSCaps{tia}}{trans-impedance amplifier}
\nAcronym{TM}{\acroSCaps{TM}}{transverse magnetic}
\nAcronym{TH4D}{\acroSCaps{th-4d}}{time domain hybrid four-dimensional}
\nAcronym{TH4D2A8PSK}{\acroSCaps{th-4d-2a8psk}}{time domain hybrid four-dimensional two-amplitude eight-phase shift keying}

\nAcronym{VCSEL}{\acroSCaps{vcsel}}{vertical-cavity surface emitting laser}
\nAcronym{VOA}{\acroSCaps{voa}}{variable optical attenuator}

\nAcronym{WDM}{\acroSCaps{wdm}}{wavelength-division multiplexing}
\nAcronym{WGA}{\acroSCaps{wga}}{weakly guiding approximation}
\nAcronym{WGN}{\acroSCaps{wgn}}{white Gaussian noise}
\nAcronym{WSS}{\acroSCaps{wss}}{wavelength selective switch}

\nAcronym{XPM}{\acroSCaps{xpm}}{cross-phase modulation}
\nAcronym{XT}{\acroSCaps{xt}}{cross-talk}

\nAcronym{3DWG}{\acroSCaps{3dwg}}{3D-waveguide} 
\usepackage[capitalise]{cleveref}

\newcommand{\LP}[1]{LP\textsubscript{#1}}

\usepackage{fancyhdr}

\fancypagestyle{firststyle}
{
  \fancyhf{}
  \fancyfoot[C]{\footnotesize \copyright~IEEE 2020}
}

\begin{document}
    \selectlanguage{english}    


    \title{\SI{1}{\tera\bit\per\second\per\lambda} Transmission Over a \SI{130}{\km} Link Consisting\\of Graded-Index \SI{50}{\micro\meter} Core Multi-Mode Fiber and 6LP Few-Mode Fiber }%


    \author{
        Menno van den Hout\textsuperscript{(1,*)}, 
        Sjoerd van der Heide\textsuperscript{(1)},
        John van Weerdenburg\textsuperscript{(1)},\\
        Marianne~Bogot-Astruc\textsuperscript{(2)},
        Adrian~Amezcua~Correa\textsuperscript{(2)}, 
        Jose~Enrique~Antonio-L\'opez\textsuperscript{(3)}, \\
        Juan~Carlos~Alvarado-Zacarias\textsuperscript{(3)}, 
        Pierre~Sillard\textsuperscript{(2)}, 
        Rodrigo~Amezcua-Correa\textsuperscript{(3)} and 
        Chigo Okonkwo\textsuperscript{(1)},
    }

    \maketitle                  


    \begin{strip}
        \begin{author_descr}

            \textsuperscript{(1)} High Capacity Optical Transmission Laboratory,
            Electro-Optical Communications Group,\\
            Eindhoven University of Technology, the Netherlands,
            \textsuperscript{(*)}{\uline{m.v.d.hout@tue.nl}}

            \textsuperscript{(2)} Prysmian Group, 644 Boulevard Est, Billy Berclau, 62092 Haisnes Cedex, France
            
            \textsuperscript{(3)} CREOL, The College of Optics and Photonics, University of Central Florida, USA

        \end{author_descr}
    \end{strip}

    \setstretch{1.1}


    \begin{strip}
        \begin{ecoc_abstract} %
            We demonstrate \SI{1}{\tera\bit\per\second\per\lambda} single-span transmission over a heterogeneous link consisting of graded-index \SI{50}{\micro\meter} core multi-mode fiber and 6LP few-mode fiber using a Kramers-Kronig receiver structure. Furthermore, the link budget increase by transmitting only three modes while employing more than three receivers is investigated.
        \end{ecoc_abstract}
    \end{strip}


%
%
%
%
%
%
%

\thispagestyle{firststyle}
    \section{Introduction}\label{sec:introduction}
    As the bandwidth requirements in various parts of optical networks rapidly approach the limits of \gls{SMF}, research towards overcoming these limitations intensifies. \Gls{SDM}, where data is transmitted over multiple modes and/or cores in a single fiber, has been shown as a promising candidate to overcome the \gls{SMF} capacity limits\cite{winzer2014optical}. It is expected that the first SDM deployments will be for applications requiring high-capacity short-reach interconnects, such as for inter-data center connections, 5G front/back-hauling and high capacity access links. However, due to the cost sensitivity of transceivers in these types of optical systems, complex and costly systems are prohibitive. The \gls{KK} receiver architecture is a potential candidate to reduce hardware complexity and its related costs. This detection method allows the recovery of the full complex field, similar to intradyne coherent detection, with a single photodiode and without the optical hybrid~\cite{AntonelliPolmuxKK}. This \gls{KK} receiver has been proven to be compatible with polarization-~\cite{AntonelliPolmuxKK,Chen2019} and space-division multiplexing~\cite{VanderHeide2018,Luis2018KK}. 
    While research towards \gls{SDM} continues at a rapid pace, it has yet to be decided which fiber technology will be dominant in future optical transmission systems. Few-mode, multi-mode, and multi-core fibers each have their own advantages and disadvantages. Separation of spatial channels in \gls{MCF} is easier compared to \gls{CCF}, \gls{FMF} or \gls{MMF}, however, \gls{MCF} manufacturing is more challenging. \gls{FMF} and \gls{MMF} on the other hand, potentially offer the highest capacity density, but heavily rely on complex \gls{DSP} to recover the spatial channels. As operators  are expected to install different fiber types in future networks, a heterogeneous optical connection might be unavoidable~\cite{Fontaine2015}.
    
    In this work, we investigate a heterogeneous single-span link of \SI{130}{\km}  comprising of \SI{73}{\km} of \SI{50}{\um} core diameter \gls{MMF} (supporting up to 45 modes) and a \SI{57}{\km} of \SI{28}{\um} core diameter 6-LP \gls{FMF} (supporting up to 10 modes). A net transmission rate of \SI{1}{\tera\bit\per\second\per\lambda} was achieved by multiplexing over 12 spatial channels (6 modes, 2 polarizations). Furthermore, we investigate the link budget increase by transmitting 3 modes while employing 4, 5, or 6 receivers. A \SI{4}{\dBm} launch power reduction was achieved by employing one additional receiver. Alternatively, an increase of  up to 0.75 in \gls{GMI} can be achieved by activating all 6 receivers at the same transmitted power.

    \section{Experimental setup}\label{sec:experimental-setup}
    \begin{figure*}[t]
        \centering
        \includegraphics[width=\linewidth]{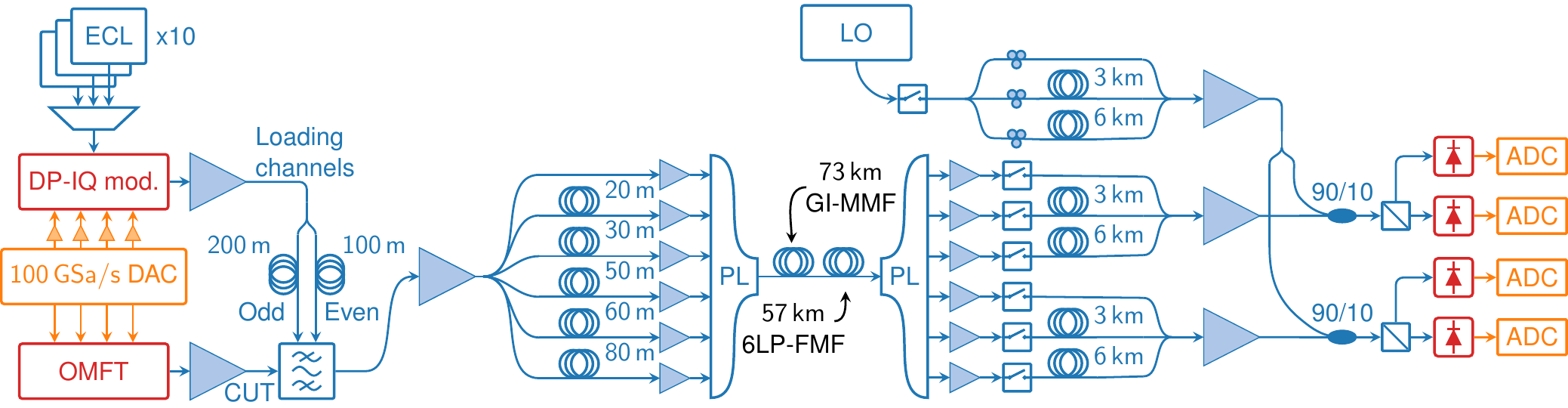}
        \vspace*{1mm}
        \caption{Experimental setup for transmission over 6 spatial modes. A \acrshort{CUT} is modulated together with 10 loading channels and is transmitted over \SI{130}{\km} of multi-mode fiber. At the receiver side, a \acrshort{TDMSDM} scheme together with \acrshort{KK} receivers is used to reduce the receiver complexity. For transmission over only 3 modes, only the first 3 EDFAs connected to the PL are enabled. 
        }
        \label{fig:setup}
    \end{figure*}
    

    The employed experimental setup is shown in \cref{fig:setup}. \Glspl{PRBS} containing \si{2^{16}} \gls{PM8QAM} symbols are generated offline at \SI{33,33}{\giga\baud}, oversampled to 3 \acrlong{SamPerSym}, pulse shaped using a \gls{RRC} filter with a roll-off of \SI{1}{\percent} and pre-emphasized for transmitter bandwidth limitations. The resulting samples are uploaded to a \SI{100}{\giga\sample\per\second} \gls{DAC}. An \gls{OMFT}, which contains RF-amplifiers, a \gls{DPIQ}, an \acrlong{ABC} and an \gls{ECL} with a \SI{100}{\kilo\hertz} linewidth is connected to the positive differential \gls{DAC} outputs to provide the \gls{CUT} at \SI{193.4}{\tera\hertz}. The negative differential \gls{DAC} outputs are amplified and used to modulate 10 tones generated by \glspl{ECL} to provide the loading channels. The loading channels are subsequently amplified and decorrelated by splitting them, delaying them using fibers of \SIlist{100; 200}{\meter}. The loading channels are multiplexed around the \gls{CUT} on a \SI{50}{\giga\hertz} \gls{WDM} grid using a \gls{OTF} which is configured such that from the \SI{100}{\m} path only the even channels pass and from the \SI{200}{\m} path only the odd channels.

    The resulting signal is amplified, split into 6 paths with lengths \SIlist{0;20;30;50;60;80}{\m} to decorrelate the tributaries. Next, the tributaries are amplified and fed into the \LP{01}, \LP{11a}, \LP{11b}, \LP{21a}, \LP{21b}, and LP\textsubscript{02} ports of a 6-mode \gls{PL}~\cite{AmadoLantern}. Disabling the \glspl{EDFA} of the three highest modes allows switching between 3-mode and 6-mode transmission. The output of the \gls{PL} is transmitted over a link consisting of \SI{73}{\km} of \SI{50}{\micro\meter} core diameter \gls{MMF}\cite{Sillard50um} and \SI{57}{\km} of 6-LP \gls{FMF}\cite{sillar6LPFMF}. In order to transition effectively from the 4-LP \gls{FMF} attached to the \gls{PL} to the large core diameter \gls{MMF}, an intermediate 6-LP \gls{FMF} is fusion spliced in between the \gls{PL} and the \gls{MMF} and funcitons as a mode adapter.

    At the receive side, the spatial channels are demultiplexed using a \gls{PL} of which the single-mode ends are fed into two \gls{TDMSDM}\cite{RoyTDMSDM} stages, which are connected to two polarization-diverse \gls{KK} receivers\cite{AntonelliPolmuxKK} with a \SI{18.5}{\giga\hertz} \gls{LO} offset. This reduces the required amount of \glspl{ADC} from 24 to 4 for this scenario. The resulting signals are digitized by a 4-channel \SI{80}{\giga\sample\per\second} \gls{ADC} with \SI{36}{\giga\hertz} of analog bandwidth, and the \gls{TDMSDM} signals are parallelized. The DC bias needed for AC-coupled \gls{KK} receivers is optimized\cite{Luis:20}, after which the \gls{KK} algorithm is performed. The residual frequency offset is estimated and compensated for, the signal is \gls{RRC} filtered, and decision-directed \gls{MIMO} equalization with in-loop blind-phase search is applied. Finally, \gls{GMI} evaluation is performed and averaged over 5 captures containing \num{241000} symbols per mode.

    \section{Results and discussion}\label{sec:results}

    \begin{figure}
        \centering
        \includegraphics[width=\linewidth]{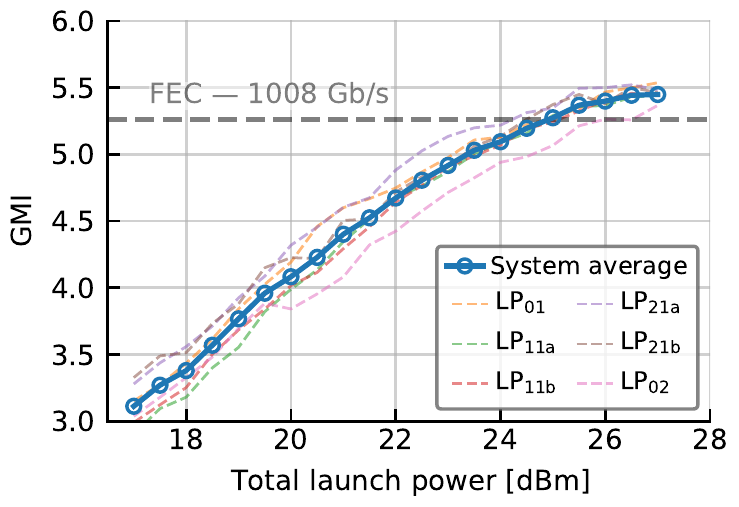}
        \caption{\gls{GMI} versus total launch power for 6 mode transmission, showing \SI{1008}{\giga\bit\per\second\per\lambda} net data rate. Dashed lines show the \gls{GMI} for individual modes.}\vspace{-2mm}
        \label{fig:6modegmi}
    \end{figure}

\begin{figure}
        \centering
        \includegraphics[width=\linewidth]{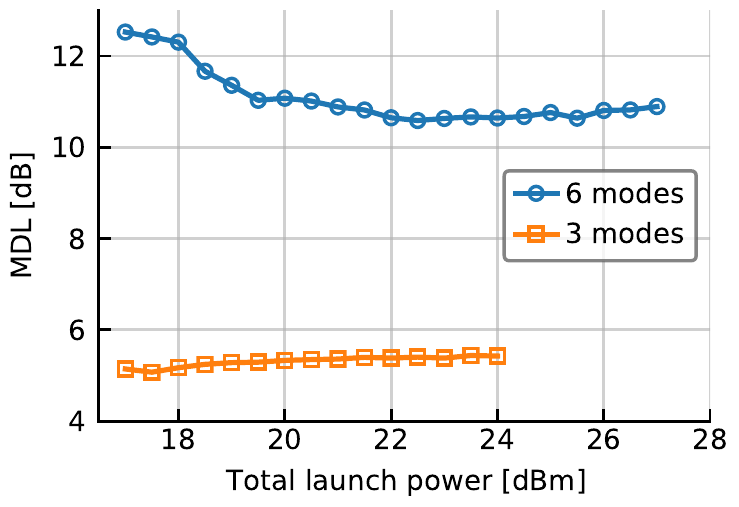}
        \caption{\Gls{MDL} derived from the equalizer taps for the 6-mode and 3-mode transmission system.}
        \label{fig:mdl}
    \end{figure}

    In \cref{fig:6modegmi}, the \gls{GMI} per mode of the \gls{CUT} is shown for the 6-mode transmission setup, indicating the system performance. In order to obtain error free transmission after \gls{FEC}, a \gls{FEC} scheme employing a spatially-coupled \gls{LDPC} combined with an outer hard-decision \gls{BCH} code is assumed, resulting in a normalized GMI limit of \num{0.8798} with a code rate of \num{0.8402} (\SI{19}{\percent} overhead)\cite{Chen2019}. Combining this with the line rate of \SI{1200}{\giga\bit\per\second\per\lambda}, a net data rate of \SI{1008}{\giga\bit\per\second\per\lambda} is achieved. 
    Furthermore, the \acrfull{MDL}, given as the ratio between the maximum and minimum of the frequency-averaged eigenvalues of the channel transfer matrix, is given in \cref{fig:mdl}. Here, the transfer matrix is obtained from the \gls{MIMO} equalizer. A \gls{MDL} of \SI{11}{\dB} can be observed for launch powers achieving the \gls{FEC} limit.

    \begin{figure}
        \centering
        \includegraphics[width=\linewidth]{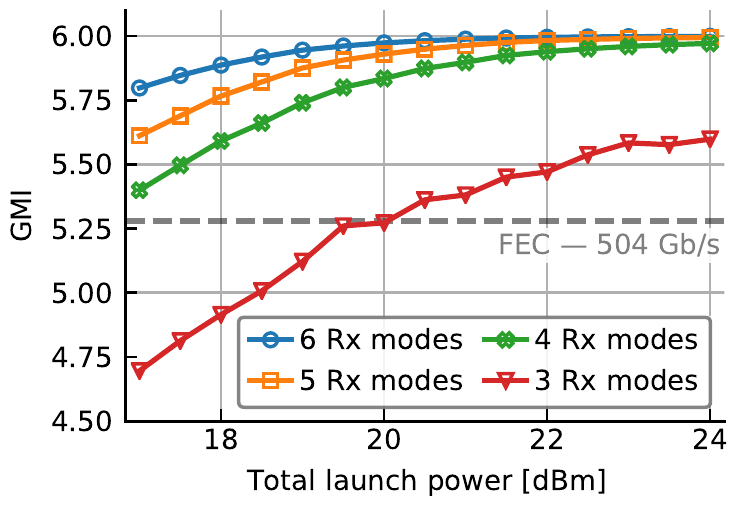}
        \caption{\gls{GMI} versus total launch power for 3 mode transmission, showing \SI{504}{\giga\bit\per\second\per\lambda} net data rate. By activating extra receivers, the link budget can be increased.}
        \label{fig:3modegmi}
    \end{figure}

    Next to the 6-mode transmission case, also a 3-mode transmission experiment is performed. At the transmitter, \glspl{EDFA} connected to the \LP{21} and \LP{02} ports of the lantern are disabled, resulting in transmitting only the \LP{01} and \LP{11} modes. For decoding at the receiver, out of the six received modes only the lower three modes are used for this transmission case. The resulting \gls{GMI} per mode is depicted by the red curve in \cref{fig:3modegmi}, achieving a line rate of \SI{600}{\giga\bit\per\second\per\lambda} and a net data rate of \SI{508}{\giga\bit\per\second\per\lambda}, assuming the same \gls{FEC} scheme as for the 6-mode transmission. From \cref{fig:mdl}, it is seen that the system \gls{MDL} is reduced compared to the 6-mode transmission system, to about \SI{5.5}{\dB}.

    \begin{figure}
        \centering
        \hspace*{-3.5mm}
        \includegraphics[width=1.05\linewidth]{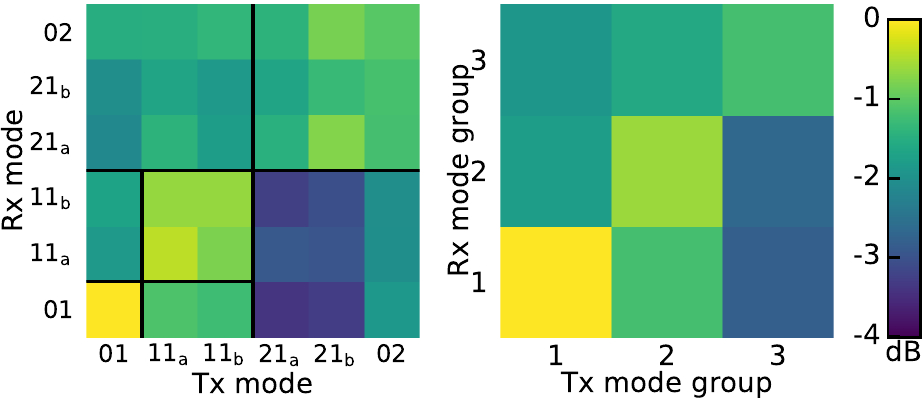}
        \caption{Transfer matrix at receiver, showing the polarization averaged (left) and the mode group averaged transmission (right), indicating strong crosstalk between the mode groups}
        \label{fig:xtalk}
    \end{figure}

    \cref{fig:xtalk} shows transfer matrix of the 6-mode system after \SI{130}{\km}. The absolute values of the time-domain taps of the \gls{MIMO} equalizer are summed per input/output combination and subsequently squared and averaged over polarizations. To obtain the mode group averaged transfer matrix shown in \cref{fig:xtalk}, the average transmission within a mode group is taken. From \cref{fig:xtalk}, it is seen that there is a relatively large coupling between the modes, especially between the transmitted \LP{01}, \LP{11a}, and \LP{11b} modes and the third mode group. This third mode group is not used for the 3-mode transmission system, so performance could be increased when also capturing information in these higher order modes.

    Therefore, the 3-mode transmission experiment is repeated, but up to 6 received modes are used at the receiver. From \cref{fig:3modegmi} it is seen that by activating extra receivers, the \gls{GMI} is increased with up to 0.75 at low launch powers due to added diversity. On the other hand, a reduction of approximately \SI{4}{\dB} in launch power can be achieved.

    \section{Conclusions}\label{sec:conclusions}
    \SI{1}{\tera\bit\per\second\per\lambda} \Gls{SDM} single-span transmission using 12 spatial channels (6 modes, 2 polarizations) over a \SI{130}{\km} link consisting of \SI{73}{\km} \SI{50}{\micro\m} core diameter \gls{MMF} and \SI{57}{\km} of 6-LP \gls{FMF} is demonstrated. Transmitting using only the first 2 mode groups yields a \SI{500}{\giga\bit\per\second\per\lambda} net data rate. By increasing the number of received modes, the link budget is increased, allowing reduction in launch power or increase in \gls{GMI}. It is expected that heterogeneous fiber links will be employed for high-capacity short-reach transmission.

    \section{Acknowledgements}\label{sec:acknowledgements}
    {%
    This work was partially funded by the Dutch NWO Gravitation Program on Research Center for Integrated Nanophotonics (Ga 024.002.033) and by the TU/e-KPN Smart Two project. Fraunhofer HHI and ID~Photonics are acknowledged for providing the Optical-Multi-Format Transmitter.
    \par}

    \printbibliography

@article{AmadoLantern,
    author = {A. M. Velazquez-Benitez and J. C. Alvarado and G. Lopez-Galmiche and J. E. Antonio-Lopez and J. Hern\' {a} ndez-Cordero and J. Sanchez-Mondragon and P. Sillard and C. M. Okonkwo and R. Amezcua-Correa},
    journal = {Opt. Lett.},
    number = {8},
    pages = {1663--1666},
    publisher = {OSA},
    title = {Six mode selective fiber optic spatial multiplexer},
    volume = {40},
    month = {3},
    year = {2015},
}

@article{Sillard50um,
    author = {Pierre Sillard and Denis Molin and Marianne Bigot-Astruc and Adrian Amezcua-Correa and Koen de Jongh and Frank Achten},
    journal = {J. Lightwave Technol.},
    number = {8},
    pages = {1672--1677},
    publisher = {OSA},
    title = {{{\SI {50} {\micro\meter}} Multimode Fibers for Mode Division Multiplexing}},
    volume = {34},
    month = {3},
    year = {2016},
}

@inproceedings{sillar6LPFMF,
    title = {{Low-DMGD 6-LP-mode fiber}},
    author = {Sillard, Pierre and Molin, Denis and Bigot-Astruc, Marianne and Maerten, H and Van Ras, Dennis and Achten, Frank},
    booktitle = {OFC 2014},
    pages = {1--3},
    year = {2014},
    organization = {IEEE}
}

@article{RoyTDMSDM,
    author = {Roy G. H. van Uden and Chigo M. Okonkwo and Haoshuo Chen and Hugo de Waardt and Antonius M. J. Koonen},
    journal = {Opt. Express},
    number = {10},
    pages = {12668--12677},
    publisher = {OSA},
    title = {Time domain multiplexed spatial division multiplexing receiver},
    volume = {22},
    month = {5},
    year = {2014},
}

@article{AntonelliPolmuxKK,
    author = {Cristian Antonelli and Antonio Mecozzi and Mark Shtaif and Xi Chen and Sethumadhavan Chandrasekhar and Peter J. Winzer},
    journal = {J. Lightwave Technol.},
    number = {24},
    pages = {5418--5424},
    publisher = {OSA},
    title = {{Polarization Multiplexing With the Kramers-Kronig Receiver}},
    volume = {35},
    month = {12},
    year = {2017},
}

@article{Luis:20,
    author = {Ruben {S. Lu\'is} and Georg Rademacher and Benjamin J. Puttnam and Cristian Antonelli and Satoshi Shinada and Hideaki Furukawa},
    journal = {Opt. Express},
    number = {3},
    pages = {4067--4075},
    publisher = {OSA},
    title = {{Simple method for optimizing the DC bias of Kramers-Kronig receivers based on AC-coupled photodetectors}},
    volume = {28},
    month = {2},
    year = {2020},
    doi = {10.1364/OE.383369},
}

@article{Chen2019,
    author = {Chen, Xi and Chandrasekhar, Sethumadhavan and Cho, Junho and Winzer, Peter},
    issn = {1094-4087},
    journal = {Optics Express},
    month = {10},
    number = {21},
    pages = {29916},
    title = {{Transmission of 30-GBd polarization-multiplexed probabilistically shaped 4096-QAM over 509-km SSMF}},
    volume = {27},
    year = {2019}
}

@inproceedings{VanderHeide2018,
    author = {van der Heide, Sjoerd and van Weerdenburg, John and Bigot-Astruc, Marianne and Amezcua-Correa, Adrian and Antonio-Lopez, Jose and Alvarado-Zacarias, Juan Carlos and de Waardt, Huug and Koenen, Ton and Sillard, Pierre and Amezcua-Correa, Rodrigo and Okonkwo, Chigo},
    booktitle = {2018 European Conference on Optical Communication (ECOC)},
    isbn = {978-1-5386-4862-9},
    month = {9},
    number = {1},
    pages = {1--3},
    publisher = {IEEE},
    title = {{Single Carrier 1 Tbit/s Mode-Multiplexed Transmission Over Graded-Index 50 um Core Multi-Mode Fiber Employing Kramers-Kronig Receivers}},
    year = {2018}
}

@inproceedings{Fontaine2015,
author = {N. K. Fontaine and T. Haramaty and R. Ryf and H. Chen and L. Miron and L. Pascar and M. Blau and B. Frenkel and L. Wang and Y. Messaddeq and S. LaRochelle and R. J. Essiambre and Y. Jung and Q. Kang and J. K. Sahu and S. U. Alam and D. J. Richardson and D. M. Marom},
booktitle = {Optical Fiber Communication Conference Post Deadline Papers},
journal = {Optical Fiber Communication Conference Post Deadline Papers},
pages = {Th5C.5},
publisher = {Optical Society of America},
title = {{Heterogeneous Space-Division Multiplexing and Joint Wavelength Switching Demonstration}},
year = {2015},
}

@INPROCEEDINGS{Luis2018KK,
  author={R. S. {Lu\'is} and G. {Rademacher} and B. J. {Puttnam} and S. {Shinada} and H. {Furukawa} and R. {Maruyama} and K. {Aikawa} and N. {Wada}},
  booktitle={2018 European Conference on Optical Communication (ECOC)}, 
  title={{A Coherent Kramers-Kronig Receiver for 3-Mode Few-Mode Fiber Transmission}}, 
  year={2018},
  volume={},
  number={},
  pages={1-3},}

@inproceedings{winzer2014optical,
  title={Optical {MIMO-SDM} system capacities},
  author={Winzer, PJ and Foschini, GJ},
  booktitle={Optical Fiber Communication Conference},
  pages = {Th1J-1},
  year={2014},
  organization={IEEE}
}

\end{document}